\documentclass[aps,prb,twocolumn,superscriptaddress,amsmath]{revtex4}

\usepackage{graphicx}
\usepackage{dcolumn}
\usepackage{amsmath}
\usepackage{bm}
\usepackage{verbatim}
\usepackage{xcolor}
\usepackage{amsbsy}

\newcommand{\be}{\begin{equation}}
\newcommand{\ee}{\end{equation}}
\newcommand{\bea}{\begin{eqnarray}}
\newcommand{\eea}{\end{eqnarray}}
\newcommand{\bean}{\begin{eqnarray*}}
\newcommand{\eean}{\end{eqnarray*}}

\newcommand{\ket}[1]{\left| #1 \right>} 
\newcommand{\bra}[1]{\left< #1 \right|}

\begin{document}

\widetext
\renewcommand{\arraystretch}{1.2}

\title{Exact special twist method for quantum Monte Carlo simulations} 

\author{Mario Dagrada}
\email[]{mario.dagrada@impmc.upmc.fr}
\affiliation{Institut de Min\'eralogie, des
Physique des Mat\'eriaux et de Cosmochimie,
Universit\'e Pierre et Marie Curie,
case 115, 4 place Jussieu, 75252, Paris cedex 05, France}

\author{Seher Karakuzu}
\email[]{skarakuzu@sissa.it}
\affiliation{International School for Advanced Studies (SISSA) 
Via Bonomea 265, 34136, Trieste, Italy}

\author{Ver\'onica Laura Vildosola}
\email[]{vildosol@tandar.cnea.gov.ar}
\affiliation{CONICET and Laboratorio Tandar,
Comisi\'on Nacional Energ\'ia Atomica,
Avenida General Paz 1499, BP 1650, Buenos Aires, Argentina  }

\author{Michele Casula}
\email[]{michele.casula@impmc.upmc.fr}
\affiliation{CNRS and Institut de Min\'eralogie, des
Physique des Mat\'eriaux et de Cosmochimie,
Universit\'e Pierre et Marie Curie,
case 115, 4 place Jussieu, 75252, Paris cedex 05, France}

\author{Sandro Sorella}
\email[]{sorella@sissa.it}
\affiliation{International School for Advanced Studies (SISSA) Via Bonomea 265,
34136, Trieste, Italy and INFM Democritos National Simulation Center, Trieste, Italy}

\date{\today}

\begin{abstract}

We present a systematic investigation of the special twist method 
introduced by Rajagopal \emph{et al.} [Phys. Rev. B \textbf{51}, 10591 (1995)] for reducing finite-size 
effects in correlated calculations of periodic extended systems with
Coulomb interactions and Fermi statistics.
We propose a procedure for finding special twist values which, at variance with previous applications of this 
method, reproduce the energy of the 
mean-field infinite-size limit solution within an adjustable (arbitrarily
small) numerical error. 
This choice of the special twist is shown to be the most accurate single-twist solution for curing 
one-body finite-size effects in correlated calculations. For these reasons we dubbed our procedure
``exact special twist'' (EST). 
EST only needs a fully converged independent-particles or mean-field calculation within the primitive cell
and a simple fit to find the special twist along a specific direction in the Brillouin zone. 
We first assess the performances of EST in a simple correlated model
such as the 3D electron gas.
Afterwards, we test its efficiency within \emph{ab initio} quantum Monte Carlo simulations of metallic
elements of increasing complexity. 
We show that EST
displays an overall good performance in reducing finite-size errors 
comparable to the widely used twist average
technique but at a much lower computational cost, since it involves the 
evaluation of just one wavefunction. We also demonstrate that the EST method shows similar performances
in the calculation of correlation functions,
such as the ionic forces for structural relaxation and the pair radial distribution function in
liquid hydrogen. 
Our conclusions point to the usefulness of EST for correlated supercell calculations; 
our method will be particularly relevant when the physical problem
under consideration requires large periodic cells.
\end{abstract}

\maketitle

\section{Introduction}

The systematic error arising from finite-size (FS) effects is a long-standing issue in computer simulations
of materials. It is well established that a poor treatment of FS errors can lead to inaccurate and unreliable
results when probing basic quantities such as total energies, structural parameters or investigating related
phenomena like phase transitions.

Among \emph{ab initio} techniques, effectively independent-electron frameworks such as Hartree-Fock (HF)
or density functional theory (DFT) can exploit the Bloch theorem for reducing FS errors.
Within a periodic system, their computational cost only depends on the size of the primitive cell and
FS effects can be controlled by averaging over a set of different boundary 
conditions (usually called \textbf{k}-points) spanning the irreducible Brillouin zone of the 
reciprocal lattice. It is in general feasible to use a large sets of \textbf{k}-points and hence
FS errors can be systematically reduced below the desired accuracy.

Despite its great success, there is a vast amount of systems where DFT has been proven insufficient.
For example, it gives unreliable predictions on strongly correlated systems such as high-temperature
superconductors, but also on moderately correlated materials requiring very accurate treatment of chemical bonds.
Phenomena developing on tiny energy scales such as the adsorption of molecules on surfaces
or proton transfer reactions in biological systems often need 
to go beyond the independent-particle framework imposed
by the DFT approach.

One of the most promising many-body methods is the set of techniques based on continuum 
quantum Monte Carlo (QMC)\cite{qmc_review1,qmc_review2}, a correlated many-body wavefunction framework. 
Due to its statistical nature, one can systematically improve the QMC precision by increasing the size of the 
statistical sample. 
QMC methods provide a truly first-principles approach to 
molecular as well as extended systems.
These features, along with the negligible parallel overhead of the main QMC algorithms, allow  
an unprecedented level of accuracy on a wide range of systems ranging from small/medium size 
molecules \cite{dagradaZundel,anthracene,polyacetylene} to strongly correlated materials as cerium\cite{CeriumQMC} and iron\cite{ironQMC}
and, recently, several high-temperature superconductors\cite{micheleFeSe,brianFeSe,lucasCuprates,spin_exchange_QMC}. 
Numerous studies have also been carried out on systems dominated by 
weak intermolecular forces such as Van der Waals interactions\cite{aromatic_rings,hay_SiO2,vdw_qmc,graph_graph_QMC} and on adsorption
phenomena \cite{adsorption_qmc,adsorption_qmc2}. 
More recently, improvements in ionic forces evaluation has led to the first 
successful attempt of fully QMC-based molecular dynamics simulations of liquid water\cite{MDhydro,MDwater}. 

While the statistical error can be systematically decreased down to the desired accuracy, the FS 
effects in a correlated framework such as QMC are considerably more delicate to deal with.
Since the electron-electron interaction is explicitly included in the \emph{ab initio} Hamiltonian solved by QMC, 
calculations
must be performed on larger simulation cells (supercells) in order to accurately take into correlations
beyond nearest-neighbors interactions.
However, even a large supercell simulation does not ensure the complete elimination of FS effects which can persist 
due to the long-range nature of the Coulomb interaction. FS effects represent therefore one of the main source of errors in 
QMC methods, often larger than the achievable statistical accuracy. 

To widen its application range, a reliable and relatively cheap method to control FS effects in QMC is certainly welcome. 
Several solutions have been proposed over the past years, which
can be divided into two main categories. The first one deals with the non-interacting part of FS errors, 
i.e. related to the kinetic energy term of the Hamiltonian and shell effects in orbital filling; this is 
in general the most important contribution to the total FS error.
The second one addresses two/many-body effects deriving from the potential energy term and from the long-range contributions
to the kinetic energy.
Examples of the first category are the twist averaged boundary conditions (TABC)\cite{ceperley_twist_average} method 
and the special \textbf{k}-points methods\cite{bloch_general,kspecial_germanium} (see Sec.\ref{sec:EST} for more details); 
the second category includes model Coulomb potential\cite{mpc_method}, KZK exchange-correlation functional\cite{kzk_funct} 
and corrections based on the random phase approximation of the electron structure factor\cite{chiesa_fse_sk}. 

In this paper, we will focus on one-body FS errors. TABC is certainly the most successful and widely used technique for tackling them within QMC simulations.
Inspired by lattice calculations \cite{gros_exact_diag,tabc_lattice}, 
it allows the many-body wavefunction to pick up a phase $\boldsymbol{\theta} = (\theta_x, \theta_y, \theta_z)$ when reaching the supercell boundaries: 
\begin{equation}
  \Psi(\mathbf{r}_1 + \mathbf{R}_s, \dots , \mathbf{r}_N) = e^{i \boldsymbol{\theta} \cdot \mathbf{R}_s} \Psi(\mathbf{r}_1, \dots , \mathbf{r}_N),
\end{equation}
where N is the number of electrons and $\{ \mathbf{R}_s \}$ denotes the supercell lattice vectors. TABC treats each twist 
independently during the simulation and therefore the resultant statistical noise when averaging over the whole set of twists is given
by $\sigma/\sqrt{N_{\mathrm{t}}}$ where $\sigma$ is the average error attained on a single twist and $N_{\mathrm{t}}$ the total number of twists.
Therefore the statistical noise is effectively reduced by performing the average and TABC requires approximately
the same amount of samples as a single-twist calculation, for a given target statistical accuracy.
This method has been proven very accurate to extrapolate to the thermodynamic (infinite-size) limit and to reduce energy 
fluctuations produced by shell filling. However, its application leads to some pitfalls. 
On one hand, it requires to keep a fixed number of fermions at each twist condition, 
i.e. standard TABC works within the canonical ensemble.
This implies that standard TABC cannot reproduce the correct thermodynamically converged independent-particle
limit of the many-body wavefunction. This issue affects the description of the Fermi surface
and it introduces a systematic small bias in kinetic energy \cite{drummond_fse}.
On the other hand, although the twist average effectively reduces statistical noise, large sets of twists
require in general a high computational burden; this is verified especially when TABC is used in combination 
with diffusion Monte Carlo (see Sec.~\ref{sec:methods}), where a relevant part of the simulation 
is spent in equilibration. 
An attempt to overcome the first drawback is the so-called grand canonical TABC (GTABC) 
method\cite{gros_gtabc,ceperley_twist_average,holzmann_HEG_fse}, where small fluctuations in the total number of 
particles $N$ are allowed for each twist condition. 
This method can be straightforwardly applied on isotropic systems with spherical Fermi surface, whereas
for realistic QMC calculations single-particle filling could be chosen according to a mean-field approach 
such as DFT. GTABC cures the kinetic energy bias introduced by the standard technique, but it leads to larger
total energy fluctuations \cite{drummond_fse}. Furthermore, it requires the wavefunction of each twist
to be optimized separately, often resulting in infeasible computational requirements for realistic systems.

In this manuscript, we propose and test a simple procedure to evaluate special 
twist values\cite{kspecial_germanium,ceperley_twist_average} for treating FS effects in correlated simulations. 
We call this procedure exact special twist method (EST). In fact, at variance with previous implementations,
the EST procedure yields the  special twist values which reproduce the exact thermodynamic limit 
of independent particles within the desired numerical
accuracy. Furthermore, we have not assumed to work with particular twist values that 
make the wavefunction real, as it is 
typically complex for generic twists.

By means of advanced variational and diffusion Monte Carlo simulations, we prove that
the EST method can eliminate most part of the energy fluctuations due to shell effects and, using a single twist, 
it shows an efficiency comparable to the TABC in thermodynamic limit extrapolation, even for systems possessing a
complicated Fermi surface.
At the same time, our method provides the correct independent-particle limit and it allows a robust optimization of the 
full (Jastrow + determinant) variational wavefunction.

The paper is organized as follows. In Sec.~\ref{sec:methods} we address in detail the QMC framework used in this paper,
we present the theoretical foundations of the special twist method and 
we outline the EST procedure. In Sec.~\ref{sec:results} we present the results.
In Sec.~\ref{sec:HEG} we assess the accuracy of our method
in a simple correlated model, the homogeneous electron gas (HEG) in three dimensions.
Then in Sec.~\ref{sec:realistic} we present realistic QMC simulations of paradigmatic metallic systems 
such as solid bcc hydrogen, the high-temperature bcc phase of lithium and the high-pressure 
$\beta$-tin structure of silicon. These systems present a different degree of complexity and difficulty 
in sampling the Fermi surface, therefore offering an exhaustive testing ground for our method.
In Subsec.~\ref{sec:energetics} we report the energetics as a function of the simulation cell size 
comparing different methodologies to reduce
FS effects. In Subsec.~\ref{sec:fs_detail} we provide a more quantitative 
assessment of the various contributions to FS errors in Li and Si.
In Subsec.~\ref{sec:derivatives} we turn our attention to correlation functions and in particular to the 
reliability of EST method in the evaluation
of QMC ionic forces. Structural relaxation on the lithium bcc cell parameters are presented
as well as pair radial distribution functions obtained from QMC molecular dynamics simulations of liquid hydrogen.
Finally we draw our conclusions in Sec.~\ref{sec:conclusions}.

\section{Methods} \label{sec:methods}

\subsection{Quantum Monte Carlo techniques} \label{sec:qmc}

Over the past years, stochastic QMC techniques are constantly gaining ground in the field of \emph{ab initio} electronic
structure calculations of solids. As mentioned in the introduction, this success is mainly due to the versatility of the 
variational wavefunction, the high accuracy attainable and the 
ability of fully exploiting the increasing computational power of modern supercomputers. 
All QMC calculations presented in this work are carried out with the \emph{TurboRVB} software package\cite{turborvb}.

The main ingredient of a reliable QMC calculation is a flexible variational ansatz.
In this manuscript, we employ the standard Jastrow-single determinant (JSD) form, used both for 
the HEG and for more realistic calculations.
Considering a system with $N$ electrons and $N_{\mathrm{at}}$ atoms, this wavefunction can be written as:
\begin{equation}
  \Psi(\mathbf{R}_{\mathrm{el}}) = \exp[-J(\mathbf{R}_{\mathrm{el}})]  \Psi^{\boldsymbol{\theta}}_{\mathrm{SD}}(\mathbf{R}_{\mathrm{el}})
  \label{eq:qmcwf}
\end{equation} 
where $ \mathbf{R}_{\mathrm{el}} = \{ \mathbf{r}_1,\dots,\mathbf{r}_N \} $ is the set of electronic positions and $\boldsymbol{\theta}$ is
the twist condition.

The Jastrow factor $J$ is the symmetric part of the wavefunction, and it is crucial for an accurate treatment of
electron correlation. Thanks to explicit two-electrons space-space correlators, it accounts in an approximate but
precise way for spatial quantum fluctuations on both charge and spin sectors. 

For HEG calculations we used: $J = \sum_{i,j}^\textrm{N} u(| \mathbf{r}_i - \mathbf{r}_j |)$ where the $u(r)$ is
a long-ranged function based on the random phase approximation (RPA). 
We refer the reader to Ref.~\onlinecite{ceperley_jas} for a rigorous derivation of this Jastrow form.

The complexity of electron interaction in realistic systems requires a 
more flexible and complete Jastrow. In order to cope with these requirements, 
we expand it on a Gaussian atom-centered basis set
$\chi_l^{\alpha}(\mathbf{r}-\mathbf{R}_{\alpha})$ where $\{ \mathbf{R}_{\alpha} \}$ ($\{ \mathbf{r} \}$) 
are the atomic (electronic) coordinates and the index $l$ spans over the whole basis set.
These orbitals do not possess any periodicity. However, when dealing with solids,
the electron-ion distances are suitably transformed in order to fulfill the periodicity of the supercell\cite{MDwater,si3}. 
The complete expression of our Jastrow is the following:

\begin{equation}
   J(\mathbf{R}_{\mathrm{el}}) = \sum_{\alpha}^{N_{\textrm{at}}} \sum_j^{N} 
   g_{\alpha}^{1b}(\mathbf{r}-\mathbf{R}_{\alpha}) + \sum_{i \neq j}^N  g^{2b}(\mathbf{r}_i,\mathbf{r}_j).
   \label{eqn:jas}
\end{equation}

The first term on the right-hand side of Eq.~\ref{eqn:jas} is a one-body factor which
accounts for electron-ion interactions:

  \begin{equation} 
    g_{\alpha}^{1b}(\mathbf{r}-\mathbf{R}_{\alpha}) = v_{\alpha}(|\mathbf{r}-\mathbf{R}_{\alpha}|) +
    \sum_l G_{\alpha}^l  \chi_l^{\alpha}(\mathbf{r}-\mathbf{R}_{\alpha})
    \label{eqn:1body}
  \end{equation}
  
where: 
 $$ v_{\alpha}(r) = Z_{\alpha} \frac{1 - e^{-\beta \sqrt[4]{(2Z_{\alpha})}r}  } { \beta \sqrt[4]{(2Z_{\alpha})} } $$
cures divergences of the electron-ion potential at coalescence points (electron-ion cusp conditions).   
The many-body $g^{2b}$ is built with the same structure as the one-body term and it reads:

    \begin{equation}
      g^{2b}(\mathbf{r},\mathbf{r}^{\prime})  =
      u(|\mathbf{r}-\mathbf{r}^{\prime}|) + 
      \sum_{{lm}}^{\alpha \beta} C_{{lm}}^{\alpha \beta} \chi_l^{\alpha}(\mathbf{r}-\mathbf{R}_{\alpha}) 
      \chi_m^{\beta}(\mathbf{r}^{\prime}-\mathbf{R}_{\beta}).
      \label{eqn:2body}
    \end{equation} 
    
This factor is designed, on one hand, to fulfill the electron-electron 
wavefunction cusps. This is achieved through a spin contaminated homogeneous term:
$$ u(r) = \frac{A}{\gamma} (1 - e^{-\gamma r}) $$
where $A=1/2$ for like spins and $A=1/4$ for unlike spins. On the other hand,
the second term on the right-hand side of Eq.~\ref{eqn:2body} 
ensures an accurate characterization of charge fluctuations by correlating single-particle
orbitals describing electrons located on different atoms.
We do not explicitly account for spin-spin correlations in Eq.~\ref{eqn:2body} as
we verified that their inclusion leads to a negligible improvement in the variational energy for all 
systems considered in this work.
The Jastrow variational parameters are therefore $\beta,\gamma$ for the homogeneous
terms, the matrix elements $G_{\alpha}^l,C_{{lm}}^{\alpha \beta}$ and the exponents of the 
Gaussian orbitals. 
For the largest simulations supercells, we set to zero the elements $C_{{lm}}^{\alpha \beta}$
connecting atoms $\alpha$ and $\beta$ whose distance $|\mathbf{R}_{\alpha}-\mathbf{R}_{\beta}|$ 
is larger than an appropriately chosen cutoff $R_{\mathrm{max}}$. 
While not affecting the final energy, this approximation considerably 
reduces the number of parameters that are effectively optimized during the minimization procedure.
This turns out in an increased stability, especially for larger systems,
by removing local minima in the energy hypersurface.

The Jastrow factor is always real-valued also in the case of non-zero twist calculations. 
Furthermore, in the TABC results presented in the next Section, we use
a common Jastrow for all twists. This assumption is physically justified by the fact that the Jastrow
is a density-density correlator function, being the (physical) electronic density a \textbf{k}-independent
quantity. The optimization of Jastrow variational parameters is based on 
the stochastic reconfiguration method which has been extensively 
described elsewhere\cite{sr1,sr2}. The atomic positions are included in the optimization procedure by treating
them on the same footing of the other variational parameters of the wavefunction.
Differently from previous implementations, in our approach the
energy derivatives with respect to the variational parameters and ionic positions are all computed 
by means of the adjoint algorithmic differentiation (AAD) introduced in Ref.~\onlinecite{sorella_aad}.
Initially devised for real-valued wavefunctions, it is straightforward to extend this technique 
to complex arithmetics. 

The antisymmetric part of our wavefunction is a complex-valued single Slater determinant represented by means of
$N/2$ molecular orbitals (MOs) $\psi_{\boldsymbol{\theta},i}^{\sigma}(\mathbf{r})$:
\begin{equation}
 \Psi_{\mathrm{SD}}^{\boldsymbol{\theta}}(\mathbf{R}_{\mathrm{el}}) = 
 \mathrm{det} \left[ \psi_{\boldsymbol{\theta},i}^{\uparrow}(\mathbf{r}_j) \right] \, \mathrm{det} \left[ \psi_{\boldsymbol{\theta},i}^{\downarrow}(\mathbf{r}_j) \right]
 \label{eq:sd}
\end{equation}
where $ 1 < i,j \leq N/2 $ and we explicitly write the dependence of the molecular orbitals
on the chosen twist value $\boldsymbol{\theta}$.
Within the \emph{TurboRVB} code implementation, it is possible to go beyond the single determinant 
representation by using a number of MOs larger than $N/2$, obtaining the so-called antisymmetrized geminal power (AGP) 
ansatz\cite{agp1,agp2}. 
However in this work we do not use the AGP extension, it is worth to mention that the methods for treating FS effects
presented here can be straightforwardly extended to the AGP. This turns out to be
of particular importance for low-energy phenomena such as high-temperature superconductivity and it
will be the subject of future publications. 
As for the Jastrow factor, the $N/2$ MOs are expanded over a periodic Gaussian basis set 
$\phi_l^{\alpha,\boldsymbol{\theta}}(\mathbf{r}-\mathbf{R}_{\alpha})$. However, in the case of the determinantal part these
functions are complex valued.
Setting $N_\mathrm{bas}$ as the total basis set dimension, the molecular orbitals read:
\begin{equation}
  \psi_{\boldsymbol{\theta},i}^{\sigma}(\mathbf{r}) = \sum_{\alpha}^{N_\mathrm{at}}
  \sum_{l}^{N_\mathrm{bas}} M_{i}^{l,\alpha} \; \phi_l^{\alpha,\boldsymbol{\theta}}(\mathbf{r}^{\sigma}-\mathbf{R}_{\alpha}) 
\end{equation}
where the optimal complex coefficients $M_{i}^{l,\alpha}$ are obtained from a DFT calculation in the local density approximation (LDA)
performed with the same setup (basis set and supercell) as the corresponding QMC calculation.
The DFT code used is built-in in the \emph{TurboRVB} package.
The orbitals of the basis set are constructed, at variance with the Jastrow factor, in a way which explicitly includes the twist dependence 
and, at the same time, satisfies the Bloch theorem for single-particle wavefunctions. 
If we denote the infinite set of 
supercell lattice vectors as $\mathbf{L}$, these orbitals read:
\begin{equation}
  \phi_l^{\alpha,\boldsymbol{\theta}}(\mathbf{r}-\mathbf{R}_{\alpha}) = \sum_{\mathbf{L}} 
  \chi_l^{\alpha}(\mathbf{r}-\mathbf{R}_{\alpha} + \mathbf{L}) e^{-i\boldsymbol{\theta} \cdot \mathbf{L}} 
  \label{eqn:basis}
\end{equation}
where $\chi_l^{\alpha}$ are the same localized Gaussian functions used for the Jastrow factor, except without 
imposing any periodicity on the coordinates. The infinite sum of Eq.~\ref{eqn:basis} is truncated above a suitable 
cutoff $\epsilon_{\mathrm{cut}}$ satisfying the following inequality:
\begin{equation}
 \zeta_l \sqrt{ L_x^2 + L_y^2 + L_z^2 } \geq \epsilon_{\mathrm{cut}} ~~~~ \mathrm{with}~l \in [1,N_{\mathrm{bas}}],
 \label{eq:cutoff}
\end{equation}
where $\zeta$ is the exponent of the localized orbital $\chi_l^{\alpha}$ and $\{ L_i \}$ are the components 
of the lattice vector $\mathbf{L}$ along the three Cartesian directions.

In the \emph{TurboRVB} implementation, we are able to relax the condition of Eq.~\ref{eq:sd} that the twist
$\boldsymbol{\theta}$ must be the same for both spin sectors. Indeed, 
this feature can
be crucial to effectively reduce FS errors in simulation of antiferromagnetic materials and 
in \emph{ab initio} high-temperature superconductivity. In fact, by setting
$\boldsymbol{\theta}^{\downarrow} = -\boldsymbol{\theta}^{\uparrow}$, we are able to preserve the 
time reversal symmetry at every supercell size. Differently from 
the standard case of equal boundaries for up/down spins, this choice also ensures the conservation
of the translational invariance of singlet electron states such as Cooper pairs. This fact
not only can improve the quality of finite-size extrapolation, as demonstrated by preliminary 
calculations\cite{parola_sorella_heis} on the Heisenberg model, but it also allows to fully exploit the translational
invariance to decrease the number of variational parameters in the determinant.
In the present work, we adopt this choice but it has no 
influence on FS extrapolation since no optimization of the determinantal part is performed. 

Except for the case of hydrogen where all electrons are included in the simulation, 
we replace core electrons with the Burkatzki, Filippi, Dolg (BFD) energy-consistent pseudopotential\cite{bfd} specifically 
designed for QMC calculations. Further details on the basis sets employed for the considered 
systems are reported in Sec.~\ref{sec:results}. 

In this article, QMC single point energies are obtained with the variational Monte Carlo (VMC) for the simplest 
systems and with the more accurate projective lattice regularized diffusion Monte Carlo\cite{lrdmc1,lrdmc2} 
(LRDMC) for the more delicate benchmarks. LRDMC, within the standard fixed-node approximation, 
allows a big improvement on the quality of the energy and correlation functions. 
The outcome of our QMC calculations for both HEG and realistic systems are detailed in Sec.~\ref{sec:results}.

\subsection{Exact special twist method} \label{sec:EST}

Within an independent-electron framework such as DFT, the Bloch theorem establishes that the thermodynamic (or infinite-size) limit of most 
physical quantities can be evaluated \emph{exactly} by performing an 
integration over the first Brillouin zone (1BZ) of the reciprocal lattice:
\begin{equation}
 f_{\infty} =\frac{\Omega}{(2\pi)^3}\int_{\Omega} d^3\mathbf{k} \, f(\mathbf{k}), 
 \label{eq:bz_int}
\end{equation}
where the function $f(\mathbf{k})$ is a periodic quantity and $\Omega$ is the volume of primitive cell.
The mean-value theorem for definite integrals ensures the existence of a special point $\mathbf{k}^*$, the so-called mean-value point, 
for which the integrand in Eq.~\ref{eq:bz_int} equals the integral, i.e. 
\begin{equation}
f(\mathbf{k}^*) = f_{\infty}. 
\label{eq:mean_value_theo}
\end{equation}
Notice that the validity of this theorem is restricted to continuous functions $f(\mathbf{k})$.

Baldereschi\cite{baldereschi} devised an analytical procedure which allows to determine, in an approximate fashion, the value of the 
mean-value point. This scheme exploits the point group symmetries of the Bravais lattice and its validity 
is restricted to slowly varying integrands $f(\mathbf{k})$. Baldereschi method gives excellent results in predicting
thermodynamically 
converged energies and other observables for insulating materials, but it fails for metallic systems where the 
integrated function in Eq.~\ref{eq:bz_int} is typically not smooth for some observables such as the total energy.
Within DFT and other mean field approaches, this issue can be cured by simply increasing the number of $\mathbf{k}$-points and approximating 
the integral in Eq.~\ref{eq:bz_int} with a sum: 
\begin{equation}
   f_{\infty} \approx \frac{1}{N_k} \sum_{i=1}^{N_k} f(\mathbf{k}_i) = \sum_{i=1}^{N_k} w_i f(\mathbf{k}_i) ~~~ \textrm{with}~~\sum_i^{N_k} w_k = 1
  \label{eq:sum_1bz}
  \vspace{0.2cm}
\end{equation}
The quality of the above approximation
can be improved by choosing a uniform mesh of $\mathbf{k}$-points in the 1BZ\cite{cohen_kpoints,mp_grid}, which is nowadays the 
standard methodology of most available DFT programs. The total number of points in the sum can be
reduced by assigning to each $\mathbf{k}$-point an appropriate weight $w_i$ determined by symmetry considerations.

QMC calculations present a different scenario. In this case, the simulation model cannot be restricted to a single primitive cell, but
a larger simulation supercell containing several primitive cells is needed due to the many-body nature of the 
\emph{ab initio} Hamiltonian. 
Refs.~\onlinecite{bloch_general,kspecial_germanium} generalized the Bloch's theorem to supercell calculations. 
Following Ref.~\onlinecite{kspecial_germanium}, the wavefunction of a 
$ n_1\times n_2 \times n_3 $ supercell (where $\{ n_i \}$ is the
number of primitive cells contained within the supercell along the three Cartesian directions) 
corresponds to $ n_1\times n_2 \times n_3 $ uniform \textbf{k}-points mesh in mean field language. Evaluating this 
wavefunction at a non-zero wave vector $\boldsymbol{\theta}_s$ is equivalent to apply an offset to 
this grid with respect to the origin at the center of the Brillouin zone. This offset is usually called ``twist'', in order
to distinguish it from the $\mathbf{k}$-points corresponding to the supercell size. 
In the independent particle limit, summing over all the twists in a given supercell is equivalent to the full \textbf{k}-point
summation in 1BZ. 

In the following, 
we present the theoretical foundations of the special twist method in supercell calculations.
At first, we set the function $f(\mathbf{k})$ in Eqs.~\ref{eq:bz_int},~\ref{eq:mean_value_theo} as the total energy,
since it is the most basic quantity to evaluate within QMC. However, this approach is general and it can be, in principle, applied also to other observables.

The basic idea behind the special twist method 
is to find a twist $\boldsymbol{\theta}_s$ which, in the limit of an independent-particle
wavefunction, reproduces the exact mean-field infinite-size energy.
The practical implementation of this idea, which we dub ``exact special twist'' (EST) method,
consists in finding an arbitrarily accurate \emph{numerical} solution to Eq.~\ref{eq:mean_value_theo}. This is carried
out within a mean-field approach which can be Hartree-Fock, DFT or any independent particle method. For most part of the calculations
presented in this manuscript we use DFT the as reference mean-field framework to solve Eq.~\ref{eq:mean_value_theo}. 
Let us consider a metallic system with $N_p$ electrons in the primitive cell, as described by an 
independent particle wavefunction such as the single determinant introduced in Eq.~\ref{eq:sd}.
The thermodynamic converged energy per electron at the DFT level reads:
\begin{eqnarray}
  \mathcal{E}_{\infty}[\rho_{\infty}] & = & \frac{1}{N_p N_k}\sum_{i=1}^{N_k} \bra{ \Psi^{\mathbf{k}_i}_{\mathrm{SD}}} 
  \mathcal{H}^{\mathrm{DFT}}[\rho_{\infty}] \ket{ \Psi^{\mathbf{k}_i}_{\mathrm{SD}} } =  \\
  & = & \frac{1}{N_p} \sum_{i=1}^{N_k} w_i \sum_{n=1} \,F(\mathcal{E}_n^{\mathbf{k}_i}[\rho_{\infty}] - \mu_{\infty}) 
  \, \mathcal{E}_n^{\mathbf{k}_i}[\rho_{\infty}], \nonumber
  \label{eq:DFT_en}
\end{eqnarray}
where the first sum goes over the $\mathbf{k}$-points in 1BZ with weights $\{ w_i \}$ and the second over 
the electronic bands $\mathcal{E}_n^{\mathbf{k}}$.
$F$ is a smearing function with Fermi distribution shape whose purpose is to smooth the electronic occupations
around the chemical potential $\mu_{\infty}$ (equal to the Fermi energy at zero temperature). Besides improving the convergence 
of the sum during the DFT self-consistent cycle, the smearing function $F$ introduces fractional electron occupations
which change the effective number of electrons considered at each \textbf{k}-point, i.e. it allows the method to work within
the grand canonical ensemble with fluctuating number of particles. The effective number of particle at each \textbf{k}-point is
determined by the chemical potential $\mu_{\infty}$.
In the case of insulators, $F$ usually takes the functional form of a Heaviside step function $H$.
In Eq.~\ref{eq:DFT_en}, we highlight that the band energies $E_n^{\mathbf{k}}$ in the sum are computed with the thermodynamic converged
electronic density $\rho_{\infty} = \frac{\Omega}{(2\pi)^3} \int_{\Omega} d^3\mathbf{k} \, \rho(\mathbf{k})$.
Notice also that $\mathcal{E}_{\infty}$ is obtained within a primitive cell simulation. 

The purpose of the EST method described here is to find a special twist value $\boldsymbol{\theta}_s$ which 
satisfies numerically Eq.~\ref{eq:mean_value_theo} up to an arbitrary accuracy. The total DFT energy for this special twist, now computed 
in a supercell with $N_s$ electrons, reads:
\begin{eqnarray}
    \mathcal{E}_{\boldsymbol{\theta}_s}[\rho_{\boldsymbol{\theta}_s}] & = & \frac{1}{N_s} \bra{ \Psi^{\boldsymbol{\theta}_s}_{\mathrm{SD}}} 
  \mathcal{H}^{\mathrm{DFT}}[\rho_{\boldsymbol{\theta}_s}] \ket{ \Psi^{\boldsymbol{\theta}_s}_{\mathrm{SD}} } = \nonumber \\
  & = & \frac{1}{N_s} \sum_{n} H(\mathcal{E}_n^{\boldsymbol{\theta}_s}-\mu_{\boldsymbol{\theta}_s}) 
  \mathcal{E}_n^{\boldsymbol{\theta}_s}[\rho_{\boldsymbol{\theta}_s}], 
  \label{eq:spec_twist}
\end{eqnarray}
where $H$ is the step function, $\rho_{\boldsymbol{\theta}_s}$ is the electronic density calculated for the special twist value, and
the chemical potential $\mu_{\boldsymbol{\theta}_s}$ is such that the effective number of electrons is equal to the true number of particles in the system. 
It is worth remarking that, in this way,
no fractional electron occupations are allowed, i.e. the energy is now computed within the canonical ensemble with fixed number of particles.
Given this formalism, finding the special twist solution to Eq.~\ref{eq:mean_value_theo} is equivalent to satisfy the 
following equality:
\begin{equation}
  \mathcal{E}_{\boldsymbol{\theta}_s}[\rho_{\boldsymbol{\theta}_s}] = \mathcal{E}_{\infty}[\rho_{\infty}].
  \label{eq:spec_twist_equal}
\end{equation}
From the above formulation, it is simple to understand that the numerical procedure to find the special twist consists in 
finding the DFT energy within the \emph{canonical} ensemble which matches, up to a certain adjustable accuracy, the \emph{grand canonical}
and thermodynamically converged energy in Eq.~\ref{eq:DFT_en}. 
Once Eq.~\ref{eq:spec_twist_equal} is solved, it is possible to build a many-body wavefunction (Eq.~\ref{eq:qmcwf}) with the chosen twist 
value which automatically fulfill the correct (at the DFT level) independent-electron limit. As we will show in Sec.~\ref{sec:results}, this
method leads to a large reduction of the one-body FS errors in QMC calculations, by keeping at the same time an affordable computational cost. 

Notice that the twist average boundary conditions method\cite{ceperley_twist_average} 
mentioned in the introduction can be viewed as the many-body equivalent of independent-particle \textbf{k}-points sampling 
in Eq.~\ref{eq:DFT_en}; indeed, all the 1BZ sampling procedures\cite{cohen_kpoints,mp_grid} exploited in DFT can be extended also to the TABC technique. 
However, the standard TABC method imposes a fixed number of particles for each twist condition $\boldsymbol{\theta}$, i.e. it works in the 
canonical ensemble only. In the independent electron framework (Eq.~\ref{eq:DFT_en}), this constraint turns out in the substitution:
$$ 
F(\mathcal{E}_n^{\boldsymbol{\theta}}-\mu_{\infty}) \rightarrow H(\mathcal{E}_n^{\boldsymbol{\theta}} - \mu_{\boldsymbol{\theta}}) 
$$ 
at each twist condition. 
It is evident that this results in a wrong independent particle limit and it leads
to an incorrect Fermi surface sampling and to a small bias in the final QMC energy\cite{drummond_fse}. This problem can be cured by
allowing the number of particles to fluctuate for each twist as implemented within the grand canonical TABC method. 
Although this method yields the correct
independent particle limit, it also gives large fluctuations in the converged energies\cite{drummond_fse} which hindered widespread use of this
technique in production QMC runs.

We turn now our attention to the practical procedure for solving numerically Eq.~\ref{eq:spec_twist_equal} in realistic QMC calculations.
The first and, at our knowledge, unique attempt to tackle this problem can be found in Refs.~\onlinecite{bloch_general,kspecial_germanium}. 
Here the authors argued that a suitable $\boldsymbol{\theta}_s$ can be chosen in the set $\{ \mathbf{G}_{\mathrm{s}}/2 \}$, where 
$\mathbf{G}_{\mathrm{s}}$ are the supercell reciprocal lattice vectors. This choice ensures that the underlined
\textbf{k}-points mesh possesses inversion symmetry, thus allowing to employ a real-valued wavefunction and avoid complex
arithmetics. The offset belonging to this set which provides the best thermodynamic limit is 
then determined via cheap DFT calculations in the LDA approximation at different supercell sizes.
The latter approach is of course approximate as the exact special twist does not necessarily falls
within the $\{ \mathbf{G}_{\mathrm{s}}/2 \}$ set. 

In this manuscript, we propose a simple but effective evolution of this procedure. 
Our methodology comprises several steps which are fully accounted in the remaining part of this Section. 

\begin{figure}
     \caption{Energy landscape of the non-interacting (NI) 2D homogeneous electrons gas with 90 particles 
     at a Wigner-Seize radius of r$_s$ = 1 a.u. as a function of the twist angle $\theta$ in $\pi/a$ units, 
     where $a$ is the cubic box parameter. Two
     representative directions in the Brillouin zone, (1,0) and (1,1), are shown. 
     The exact value of the NI energy is 
     represented by the straight line. The energy surface presents some cusps which are due to 
     discontinuous changes in the occupations of the electronic states.
     We notice that several twist conditions in both directions match the 
     value of $\mathcal{E}_{\infty}^{\mathrm{HEG}}$. We verified that the choice of both the direction and the 
     specific special twist is irrelevant for the final outcome. 
     } \label{fig:heg_EST}
     \includegraphics[width=\columnwidth]{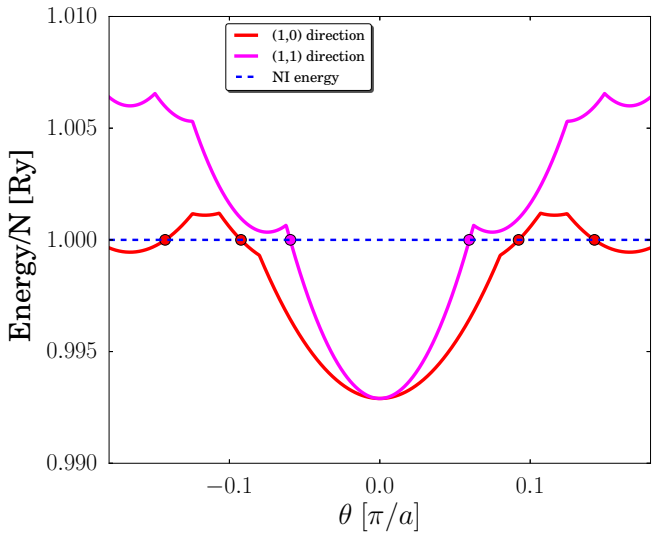}
\end{figure}

\begin{enumerate}
\item At first we determine the thermodynamic 
converged energy $\mathcal{E}_{\infty}$ (Eq.~\ref{eq:DFT_en}) within an independent-particle 
or mean-field approaches. 
For 3D-HEG calculations (Subsec.~\ref{sec:HEG}), this 
reference energy is the non interacting (NI) energy: $\mathcal{E}_{\infty}^{\mathrm{HEG}} = \sum_\mathbf{k} \mathbf{k}^2/2 \simeq \frac{2.21}{r_s^2}$ 
whose value is completely controlled by the Wigner-Seize radius $r_s$, i.e. by the electronic density.
For realistic QMC runs (Subsec.~\ref{sec:realistic}), we evaluate it at the DFT-LDA level using a fully 
converged \textbf{k}-points mesh\cite{mp_grid} and with the same basis set as QMC. 
All these calculations are carried out in the primitive cell with a negligible computational cost 
as compared to QMC.

\item The second step consists in the numerical solution of Eq.~\ref{eq:spec_twist_equal}, at given fixed number of particles.
In the case of HEG, we select several high symmetry
directions and we scan the reciprocal space along these directions in order to find the value of the twist $\boldsymbol{\theta}_s$ giving the
exact non-interacting energy $\mathcal{E}_{\infty}^{\mathrm{HEG}}$. In Fig.~\ref{fig:heg_EST} the simple case of the two dimensional
electron gas with r$_s$ = 1 a.u. is shown. We notice that
several twist values satisfy the condition in Eq.~\ref{eq:spec_twist_equal}. 
We verified that the final result is independent both on the direction and on the special
twist selected. The same procedure can be straightforwardly applied to the three dimensional electron gas 
reported in Subsec.~\ref{sec:HEG}.
Similarly, in the case of realistic QMC calculations we pick a direction in the first Brillouin zone
(in general along a diagonal, thus characterized by just one parameter) and we scan the DFT-LDA band structure 
by computing energies at each twist on a uniform grid along the chosen direction. 
These runs must be performed within the same supercell used for QMC (Eq.~\ref{eq:spec_twist}).
We select the value which reproduces the thermodynamic limit within a range smaller than the accuracy 
required by QMC. In particular, for calculations on metallic systems presented in Subsec.~\ref{sec:realistic}, we
choose an accuracy of 0.005 eV/atom in determining the special twist.
Notice that the selected twist will likely require a complex-valued determinantal part in the QMC variational ansatz. 
In contrast with Refs.~\onlinecite{bloch_general,kspecial_germanium}, this is the case for all calculations presented in Subsec.~\ref{sec:realistic}.

\item Once the special twist value $\boldsymbol{\theta}_s$ is found, we perform a final DFT-LDA supercell calculation
with the selected twist. The resulting wavefunction is used as determinantal part of the total QMC ansatz (Eq.~\ref{eq:qmcwf}).

\item Given the JSD wavefunction built in the previous steps, we carry out the Jastrow optimization with the 
stochastic reconfiguration technique (see Subsec.~\ref{sec:qmc} for technical details). Both the linear coefficients and the 
Gaussian exponents of the Jastrow are optimized. In the case of bcc-Li structure in Subsec.~\ref{sec:realistic} 
we also test the effectiveness of special twist method in predicting structural properties by performing
full QMC crystal cell relaxation.

\item The final QMC energy is evaluated with the variational and, in selected cases, 
diffusion Monte Carlo schemes using the JSD ansatz. The procedure is repeated at different supercell sizes in order 
to perform an extrapolation to the infinite-size limit. 

\end{enumerate}

In conclusion, our procedure for finding special twist values is cheap and its numerical accuracy can be adjusted depending
on the considered physical problem. The way we select the EST ensures the correct independent-particle
limit for the many-body QMC wavefunction without relying on grand canonical ensemble formalism.
Besides these features, our approach also possesses the advantages of employing only a single twist: the computational 
cost of a EST calculation is about twice the cost of a simple gamma point simulation due to the complex arithmetics needed by the 
wavefunction evaluation.

\section{Results} \label{sec:results}

This Section presents the analysis of finite-size effects for several systems by means of the EST method 
introduced in Sec.~\ref{sec:EST}. Sec.~\ref{sec:HEG} addresses the homogeneous electron gas in 3 dimensions, a correlated model 
for metallic systems widely used as a benchmark for FS correction techniques.
In Subsec.~\ref{sec:realistic} we focus on paradigmatic \emph{ab initio} metallic systems 
and we compare EST with other FS methodologies.
In particular, we test it against the simple periodic boundary conditions (PBC),
the standard twist average boundary conditions (TABC) technique\cite{ceperley_twist_average} and a 
different special twist determined analytically with the procedure introduced in Ref.~\onlinecite{baldereschi}, which we dub 
``Baldereschi point''. At first we analyze the effectiveness of EST method in extrapolating the total
energy to the infinite-size limit. 
If not otherwise specified, 
the energies are corrected 
for many-body finite-size effects using the KZK energy functional introduced 
in Ref.~\onlinecite{kzk_funct}. DFT(KZK) is performed with the built-in DFT code 
of the \emph{TurboRVB} package. The two-body corrections\cite{si3} we apply to the total
QMC energy read: 
\begin{equation}
\epsilon_\mathrm{2b}^{\mathrm{KZK}} = \mathcal{E}_\mathrm{LDA}^{N_s} - \mathcal{E}_\mathrm{KZK}^{N_s}, 
\label{eq:kzk}
\end{equation}
where both LDA and KZK energies are computed within the same $N_s$ electrons supercell and the same twist(s) condition(s)
as the corresponding QMC calculation.
We verified that the application of the corrections based on the RPA electronic
structure factor \cite{chiesa_fse_sk} leads to very similar results. A brief discussion on the evaluation of many-body
errors directly at QMC level is also presented.

The last part of Subsec.~\ref{sec:realistic} analyzes FS effects on correlation functions using the EST method. In 
particular, we report the results on the bcc-Li lattice constant evaluated with a zero temperature structural relaxation
of QMC ionic forces. Finally, we present some benchmark calculations on radial pair distribution functions
extracted from QMC based molecular dynamics simulations of high-pressure high-temperature hydrogen\cite{MDhydro,MDhydro2}.

\subsection{3D homogeneous electron gas} \label{sec:HEG}

The homogeneous electron gas in 3 dimensions (3D-HEG)
is certainly the most studied model for correlated metallic systems. Its importance is not limited
at the model level, but it also constitutes the basis for building the local density approximation
routinely employed in density functional theory\cite{ceperley_lda}.

\begin{figure}
     \caption{ Comparison of different method for alleviating FS effects in the 3D electron gas. The 
     arrow indicates the infinite-size limit as presented in Ref.~\onlinecite{ceperley_twist_average}. The solid
     line on EST results is obtained with a simple quadratic polynomial fit.} \label{fig:heg}
     \includegraphics[width=\columnwidth]{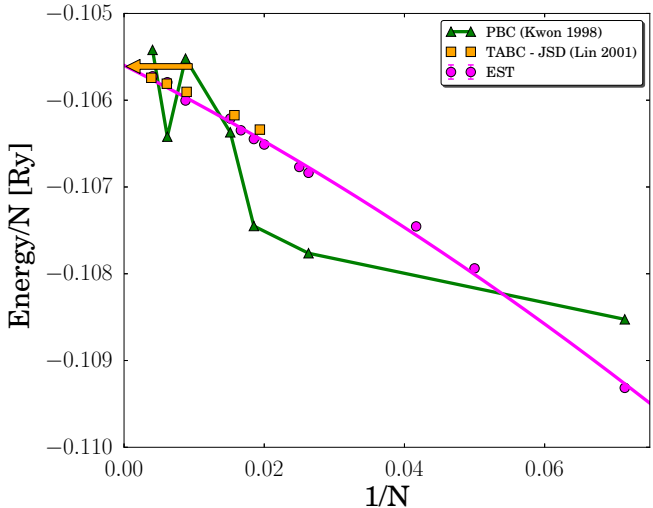}
\end{figure}

In this work, we simulated the 3D-HEG at an electronic density corresponding to a
Wigner-Seitz radius of r$_s$ = 10 a.u. This value has been used in several published works\cite{ceperley_twist_average,chiesa_fse_sk} 
carried out with the TABC method and it is therefore convenient for the sake of comparison.
In Fig.~\ref{fig:heg} we present the FS size extrapolation of HEG total energy per electron
as a function of the inverse number of particles.  
We analyze the performance of EST by comparing it with simple PBC\cite{backflow_heg} and with
TABC calculations\cite{ceperley_twist_average} both carried out with the same Slater-Jastrow trial
wavefunction as ours. The arrow in Fig.~\ref{fig:heg} indicates the infinite-size
limit as presented in the original manuscript in Ref.~\onlinecite{ceperley_twist_average}. 
We did not apply any many-body FS correction to our results. 
 
We can immediately notice that both EST and TABC are effective in suppressing shell fluctuations 
as the number of particles grows. Both methods yield a very smooth curve and we can easily extrapolate our
EST results to the infinite-size limit using a quadratic polynomial fit. 
Furthermore, we can directly match the two thermodynamic limits of the variational energy. 
TABC and EST are in agreement within the statistical error bar $ \sigma \sim 2 \times 10^{-5} Ry$, 
the extrapolated energy per particle of EST being -0.10558(2) $Ry/N$ and 
-0.10561(5) $Ry/N$ the corresponding TABC one.

At variance with the other methods, the shell fluctuations in the PBC energies are too large 
to perform any extrapolation, as expected. This issue is apparent also
if one considers the infinite-size estimate of -0.10549(2) $Ry/N$ reported in Ref.~\onlinecite{backflow_heg}.
We notice that it still displays
a discrepancy of the order of $\sim 5 \sigma$ with respect to EST and TABC infinite-size limits. 
This disagreement is likely due to residual one-body error dependence which cannot 
be suppressed despite the very large number of particles employed for this simulation. 

\begin{figure}
     \caption{ VMC extrapolation to the infinite-size limit on solid bcc structure of hydrogen. Results are corrected
     for two-body errors with the KZK method\cite{kzk_funct}. The $x$ axis reports 
     the inverse number of atoms in the supercell. 
     In the inset we zoom the results for the largest
     supercells in order to appreciate the thermodynamic limit convergence. 
     The $x$ axis of the inset reports the actual number of hydrogen atoms 
     present in the supercell.} \label{fig:hydrogen}
     \includegraphics[width=\columnwidth]{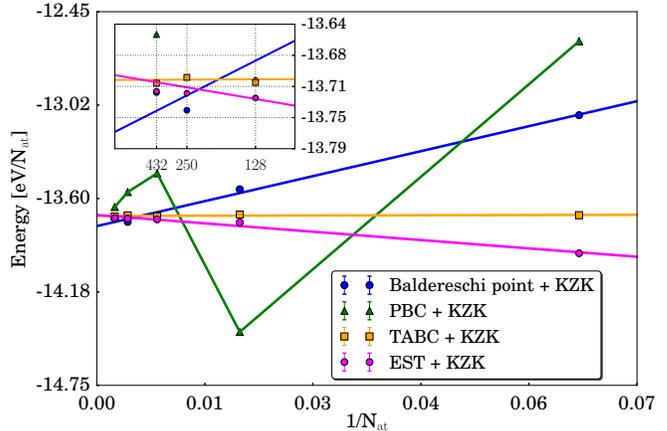}
\end{figure} 

\begin{figure*}[!htp]
\centering
\caption { Energy extrapolation on Li in the high temperature bcc
  phase. We show VMC results (panel a) and 
           LRDMC energies (panel b). The energies are compared
           with different techniques. Two-body corrections are applied with the KZK functional approach. 
           In the inset, a zoom of the results for the largest supercells is shown.
           The $x$ axis reports the inverse number of atoms 
           in the supercell for the main plot and the number of atoms itself for the inset. } \label{fig:lithium_fse}
\begin{tabular}{cc} 
      \includegraphics[width=\columnwidth]{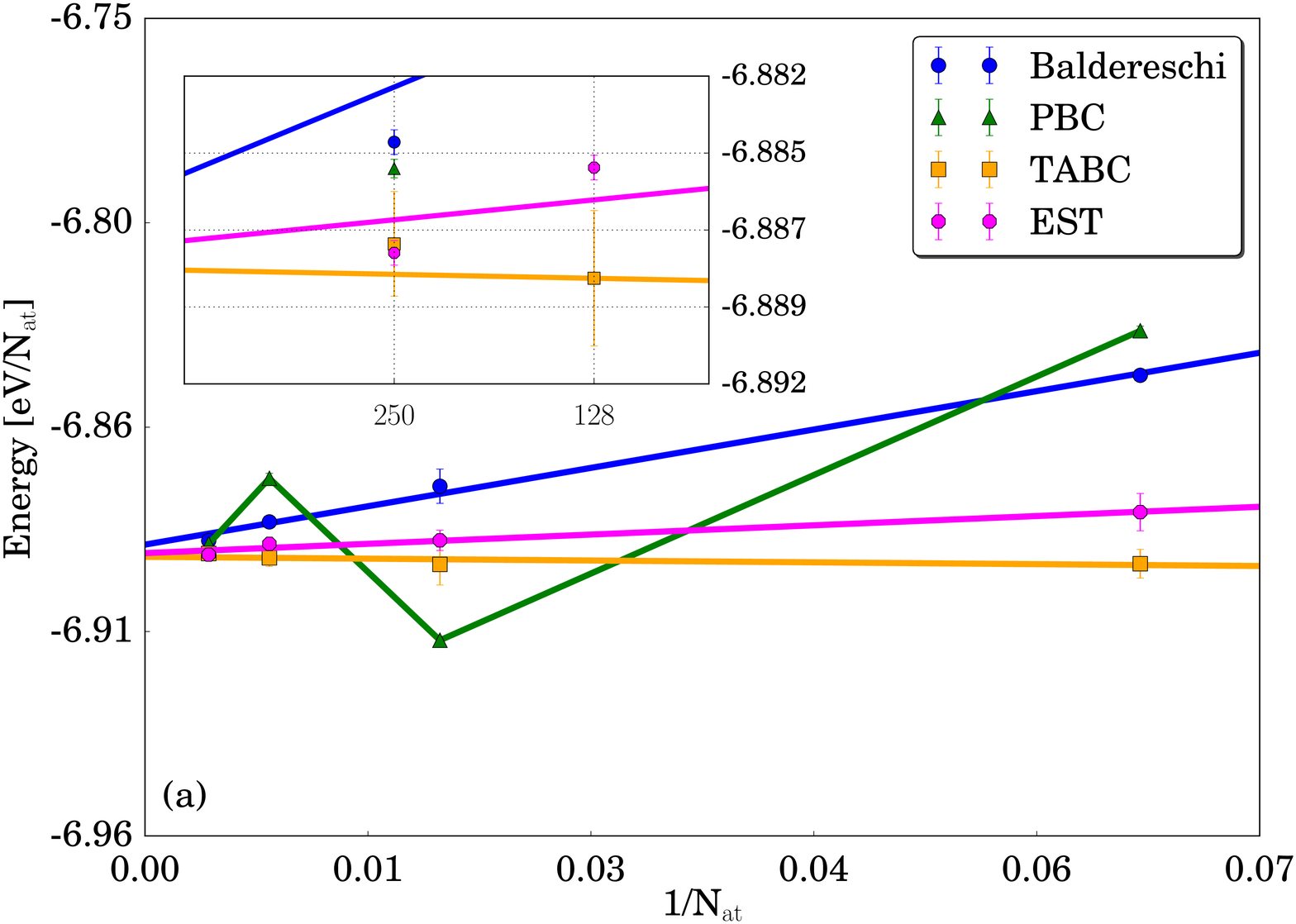} &
      \includegraphics[width=\columnwidth]{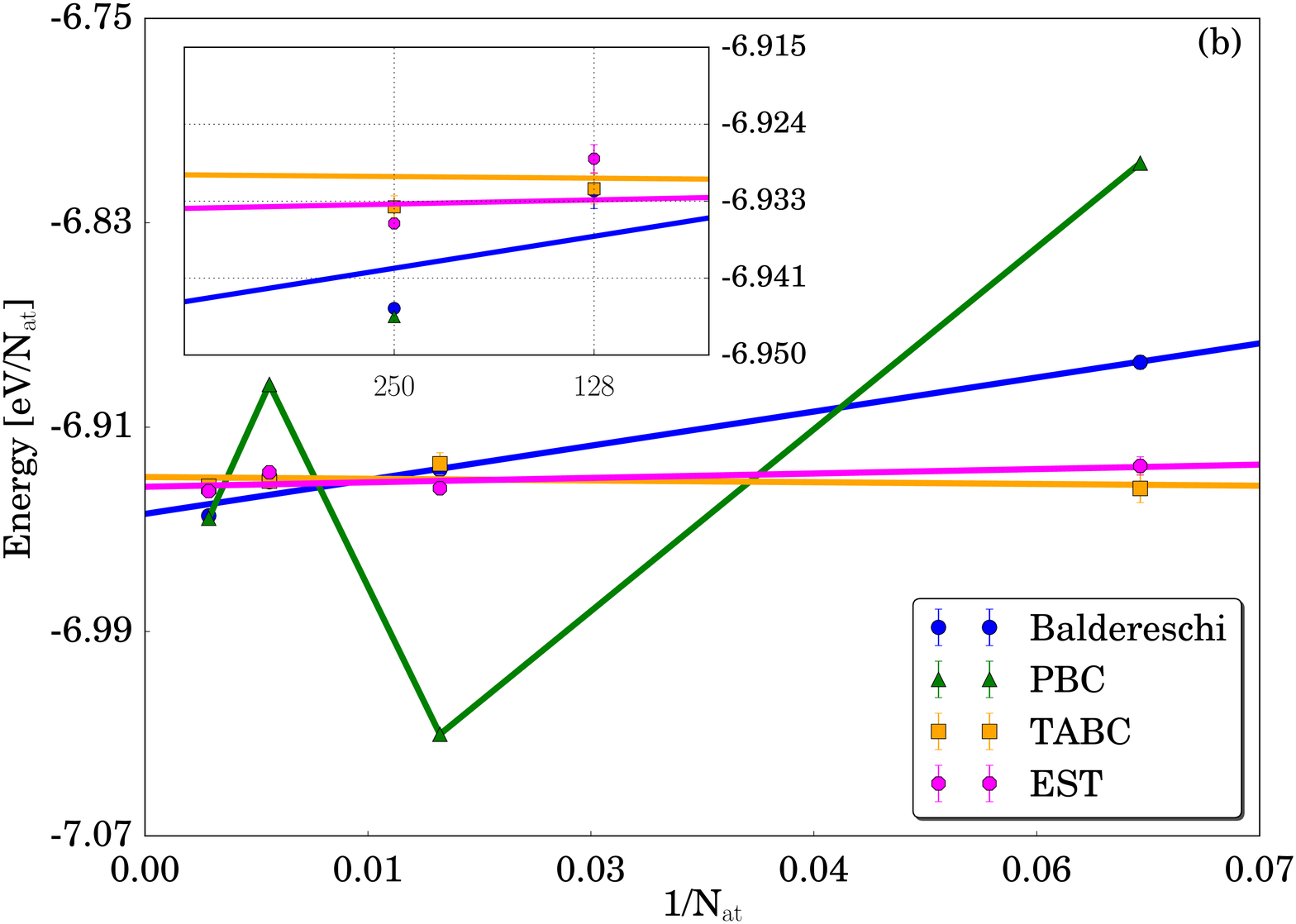} \\
\end{tabular}
\end{figure*}

\subsection{Realistic systems} \label{sec:realistic}

\subsubsection{Total energy}
\label{sec:energetics}

In this Section we analyze the FS effects on the energetics of 
three paradigmatic metallic systems with increasing degree of complexity. 
We believe they constitute an exhaustive testing ground
for the EST method and they pave the way for applying EST to more complex compounds.
For TABC calculations we choose a uniform Monkhorst-Pack\cite{mp_grid} mesh offsetted from the $\Gamma$ point 
of the supercell Brillouin zone. The number of independent twist conditions is reduced using the point group symmetry 
operations of the supercell lattice. 
In order to ensure convergence, the mesh size is varied at each supercell size 
such that the corresponding number of inequivalent atoms is kept constant and appropriately large. The
size of the twist mesh is reported on the top $x$ axis of each plot.

The first metallic system we address is solid hydrogen in the bcc structure. Despite being the simplest
element in the periodic table, hydrogen displays very intriguing properties and its phase diagram under pressure 
is far from being completely understood. In particular, in the region up to $\sim 300$ GPa, solid hydrogen
undergoes numerous phase transitions displaying exotic quantum properties which are 
not well characterized yet either experimentally or 
theoretically\cite{revHydrogen,drummond_hydro,azadi_hydro}.
It is well established that FS effects represent an important source of error in 
many-body simulations and the size of the simulation supercell is crucial for obtaining accurate correlation
functions in molecular dynamics simulations of liquid hydrogen, as we will show in Sec.~\ref{sec:derivatives}. 
Here we study the bcc structure of solid hydrogen which has not been observed yet in nature, but 
it is one of the candidate structure for the high-pressure atomic phase due 
to its dense packing of the atoms\cite{louie_book}.

We use a primitive Gaussian basis set of $[J]2s[D]2s$, 
where J refers to the Jastrow and D to the determinantal part. The exponents of the determinant are taken 
from a previous fully optimized calculation\cite{MDhydro2}. Despite its small size, this basis has been proven 
accurate in describing both energetics and the most important correlation functions\cite{MDhydro,MDhydro2} of hydrogen.
FS extrapolation at a VMC level is presented in Fig.~\ref{fig:hydrogen}. As apparent, both TABC and EST show a very smooth behavior toward the thermodynamic limit, indicating that most part of shell fluctuations have been eliminated. 

By performing extrapolation to infinite size, we obtain -13.7117(38) $eV$/atom for TABC and -13.7197(43) $eV$/atom using our EST method. These two FS correction methods are therefore in excellent agreement for a simple, but relevant system as hydrogen. 
If the Baldereschi point\cite{baldereschi} is used to offset the twist grid, the energy fluctuations are mostly
suppressed, but the extrapolation procedure does not yield a satisfactory result (see the inset of Fig.~\ref{fig:hydrogen}).

\begin{figure}
     \caption{ Fermi surface contour plots along the $z$ axis for bcc-Li (left panel) and $\beta$-tin Si (right panel). 
     The contour has been taken at k$_z$ = 0. The calculations have been performed with the software package Wannier90\cite{wannier}
     based on DFT(LDA) results obtained with the QuantumESPRESSO\cite{qe} program. } \label{fig:fermi}
     \includegraphics[width=\columnwidth]{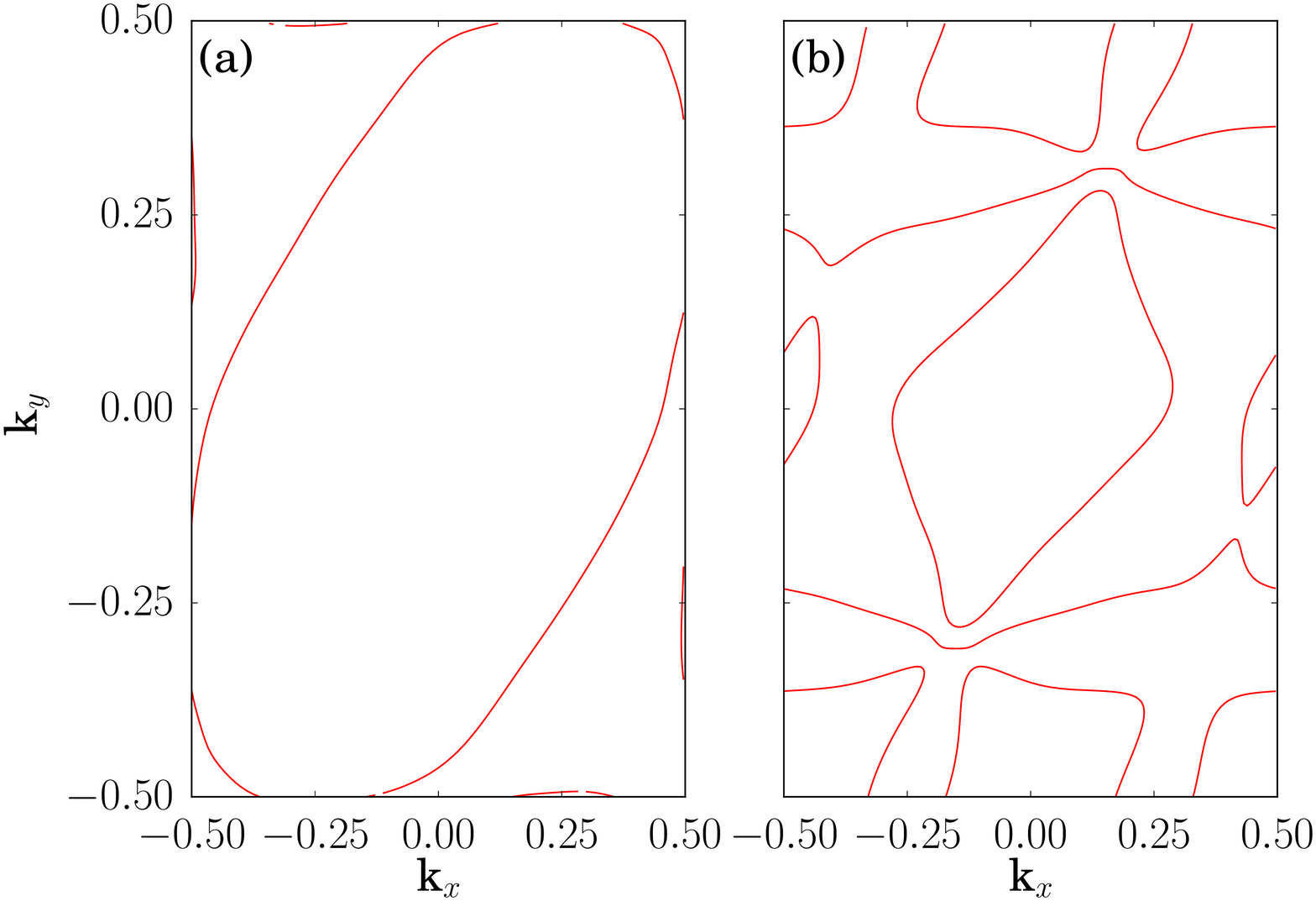}
\end{figure}

\begin{figure*}[!htp]
\centering
\caption { Finite-size extrapolation with VMC (panel a) and LRDMC (panel b) methods on Si in 
     the high pressure $\beta$-tin phase. Again a comparison is 
     shown among various FS correction methods and the inset reports a zoomed view close to the thermodynamic limit. 
     Two-body errors are corrected with KZK method. The $x$ axis the inverse number of atoms for 
     the main plot and the number of atom itself for the inset. Except for PBC, we perform a linear fit on the results.
     We notice that, despite yielding a smooth curve, the EST energy in 
     the 16 atoms supercell (second point from the right) displays a shift towards higher energies.
     This is likely due to a particularly sharp Fermi surface for this specific supercell; we verified 
     that, at DFT level, the use of a rotated supercell allows to partially suppress this shift. For consistency,
     we keep the result in the plot but we remove it from the fitting
     procedure. } 
\label{fig:silicon}
\begin{tabular}{cc} 
      \includegraphics[width=\columnwidth]{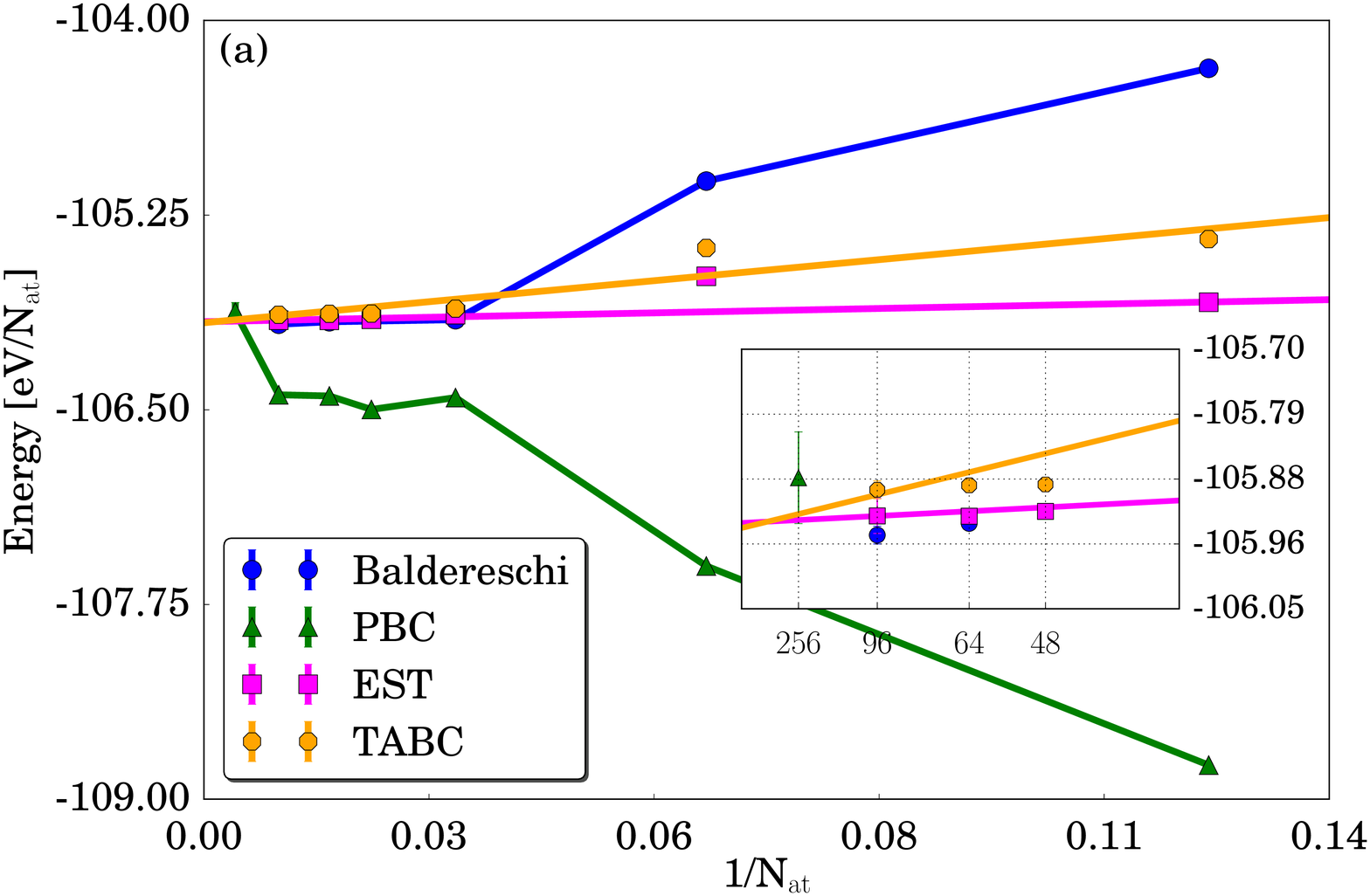} &
      \includegraphics[width=\columnwidth]{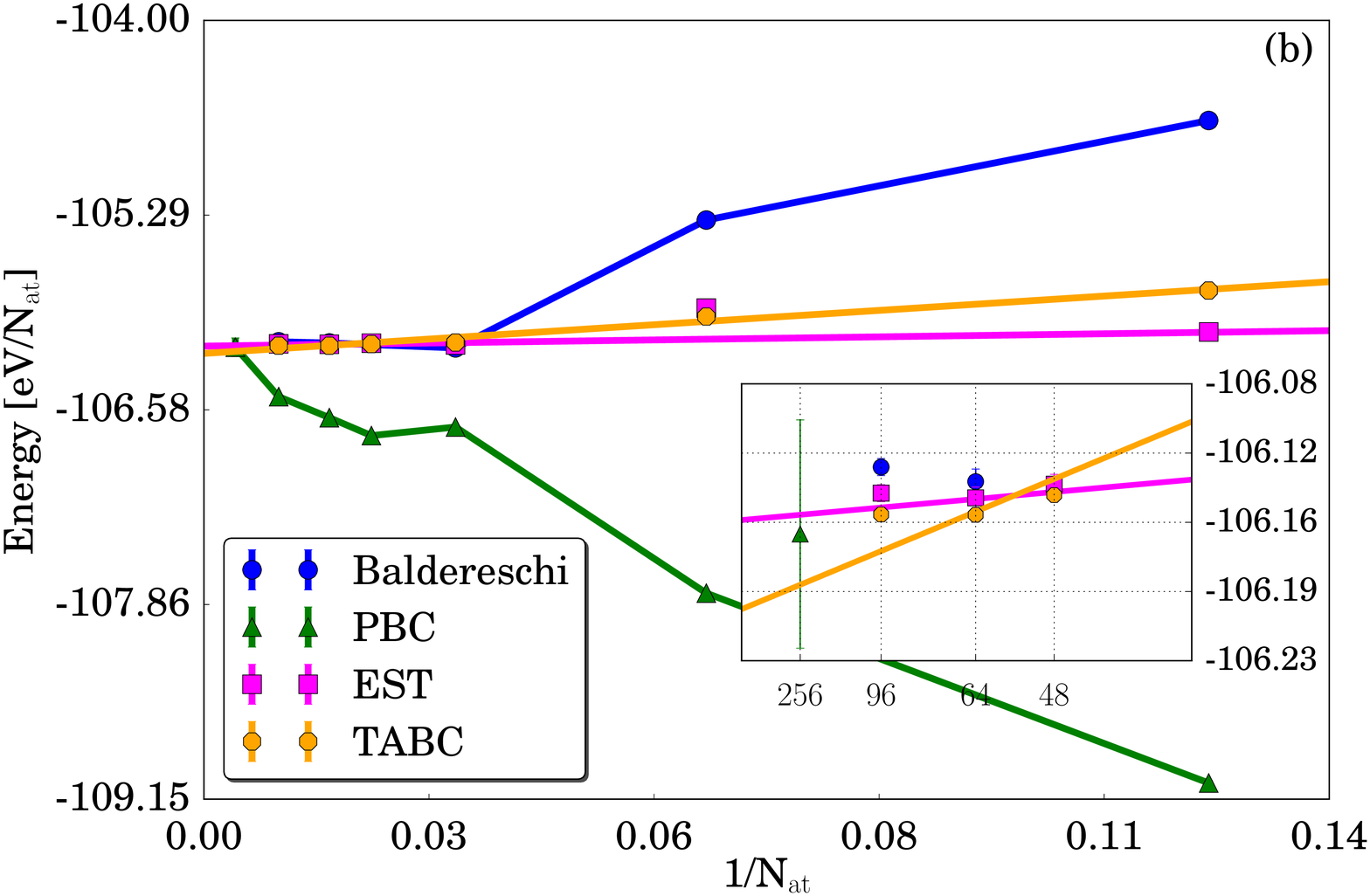} \\
\end{tabular}
\end{figure*}

We turn now our attention to metallic bcc lithium.
Bulk Li has been the subject of intense studies due to the emergence of exotic quantum states, including
superconductivity\cite{li_sc,li_sc2}, in its phase diagram under pressure and also to its extensive 
application in battery development.
Previous QMC investigations\cite{li_nanoclusters,mitas_lithium} provided very accurate results and, at the 
same time, they established the important role of FS effects in determining the converged ground 
state energy\cite{mitas_lithium,azadi_fse_study}.
For treating this system the localized basis set used in this work is $[J]2s2p[D]4s4p$ and the 1s core 
electron is replaced with a BFD pseudopotential\cite{bfd}.
A comparison among several FS methods is presented in Fig.~\ref{fig:lithium_fse}a for VMC and Fig.~\ref{fig:lithium_fse}b for 
LRDMC energies. 
Both TABC and EST methods, in combination with the KZK corrections, ensure an almost complete suppression of energy 
fluctuations and they provide a well converged result already for the $54$ atoms supercell, 
at variance with the Baldereschi point which displays a much slower convergence with supercell size.
In order to fully appreciate the convergence to the thermodynamic limit, a zoom on the largest systems
is reported in the inset. The final extrapolated results of TABC and EST are in agreement up to $0.002$ eV/atom
for both VMC and LRDMC. This value is of the order of the attained statistical error.  
These results demonstrate that TABC and EST provide a similar performances 
in controlling FS effects in this system, although the former displays a slightly 
flatter curve. 

The last system we address for benchmarking our method is the high pressure $\beta$-tin structure of silicon. 
Upon application of a pressure around 12 GPa, Si displays a structural phase transition 
from the semiconductor diamond phase to a $\beta$-tin metallic phase. The transition develops on a very narrow
energy scale\cite{si_exp} and standard DFT techniques yield unsatisfactory and functional-dependent results. 
Due to its sensitivity, this phenomenon is a perfect ground for benchmarking advanced first principles
methods such as QMC and it has been extensively studied\cite{si1,si2,si3,si4,si5}. 
The tiny energy scale ($\sim 0.05$ eV/atom) to be probed in order
to spot the correct transition pressure, requires a very accurate control of finite-size effects \cite{si3}.
Metallic Si offers a perfect playground for testing the reliability of EST method when tackling systems with 
complex and discontinuous Fermi surfaces. 
To be more explicit, in Fig.~\ref{fig:fermi} we show a comparison of the LDA Fermi surface 
between bcc-Li and $\beta$-tin Si,
where the contour has been taken along the $z$-axis at $\mathbf{k}_z = 0$. We notice that Li (Fig.~\ref{fig:fermi}a) displays 
practically no features except a large electron pocket centered at the $\Gamma$ point. $\beta$-tin 
Si (Fig.~\ref{fig:fermi}b) is instead considerably more challenging, 
in particular as a result of the small electron pockets present at 
the Brillouin zone borders. 
As already presented in the Li case, we report FS extrapolation in Si with KZK corrections for curing
two-body FS effects.
The final results obtained with the VMC and LRDMC methods are shown in Fig.~\ref{fig:silicon}a and 
Fig.~\ref{fig:silicon}b respectively.
The EST method gives excellent results, comparable to the more expensive TABC in eliminating 
shell filling effects. In this particular case, the energy curve obtained with EST method is 
flatter than the TABC one. 
However, we notice that the 
energy of the 16 atoms supercell (second point from the right) is shifted towards higher energies.
We believe that this issue is related to the particularly poor sampling of the Fermi surface 
for that set of $\mathbf{k}$-points. 
This fact together with the relatively small number
of atoms in the supercell -- which corresponds to a coarse $\mathbf{k}$-points mesh --
may give rise to the shift observed in EST results.
We verified, at DFT level, that this shift can be partially recovered by using 
a rotated supercell with lower symmetry. In Fig.~\ref{fig:silicon} we do not 
include this point in the extrapolation procedure.

The infinite-size extrapolation of TABC and EST are in agreement up to $0.005$ eV/atom for the VMC method, 
a value below the attained statistical error. A larger difference is apparent when using the 
LRDMC; beyod-leading-order extrapolation might be more appropriate in this case. However, when the largest supercell (96 atoms for 
EST, Baldereschi and TABC, 256 atoms at $\Gamma$ point) is used, all results, independently from the method used, are converged 
up $0.01$ eV/atom for all techniques, an accuracy sufficient for obtaining a correct transition pressure\cite{si3}. 

\subsubsection{Comparison of errors in the EST method}
\label{sec:fs_detail}

\begin{table}[!htp]
\begin{center}
\begin{tabular}{c|ccc} 
        
    & $\epsilon_{\text{2b}}^{\mathrm{KZK}}$ & $\epsilon_{\text{2b}}^{\mathrm{VMC}}$ & $\epsilon_{\text{1b}}^{\mathrm{VMC}}$ \\
   \hline
   $\#$ \textrm{of atoms} & \multicolumn{3}{c}{bcc-Li} \\
   \hline
    16      & 0.1958 & 0.198(7) & -0.013(6)   \\
    54      & 0.0378 & 0.061(5) & -0.006(5)   \\
    128     & 0.0245 & 0.026(5) & -0.003(2)   \\
    250     & 0.0125 & 0.013(5) &  0.0002(17) \\
   \hline
   & \multicolumn{3}{c}{$\beta$-tin Si} \\
   \hline
    16     & 0.3545 & 0.185(6) &  0.026(2) \\
    48     & 0.1181 & 0.096(7) &  0.036(7) \\
    64     & 0.0886 & 0.067(6) &  0.042(6) \\
    96     & 0.0574 & 0.042(8) &  0.035(2)  \\
  \hline  
\end{tabular}
\caption { Comparison of many-body FS effects estimated with different methods for both bcc-Li and 
           $\beta$-tin Si . All presented results are in eV per 
           atom. Many-body errors are compared between the KZK and VMC corrections, both detailed in the 
           text. We notice an overall good agreement between the two estimations for both Li and Si, except
           the case of the 16 atoms Si supercell which displayed similar issues also in the energy 
           extrapolation (see Fig.~\ref{fig:silicon}).
           } \label{tab:2b_errors}
           \vspace{0.1cm}
           \hspace{-0.2cm}
\end{center}
\end{table} 

In this Section we present a more quantitative discussion on the impact of FS effects 
in the special twist approach.

By definition (Eq.~\ref{eq:spec_twist_equal}), the EST method cancels out all FS errors derived from
the one-body contribution at mean-field level. 
However, if one switches electron correlation on in QMC,
the Fermi surface can vary from the single-particle estimation, thus reintroducing
some one-body finite-size effects also in the case of EST. Their size can be estimated 
via the TABC technique which ensures a denser sampling of the Fermi
surface. Thus,
we provide an estimation of this residual contribution to the one-body FS errors 
directly within QMC. 
For this estimate VMC is the method of choice as it likely provides an outcome
similar to LRDMC concerning FS effects and, since it is much cheaper, it can be used 
in production runs for correcting the LRDMC energy results.
The residual one-body error at VMC level reads:
\begin{equation}
 \epsilon_{\text{1b}}^{\mathrm{VMC}} = \mathcal{E}_\mathrm{VMC}^{\mathrm{TABC},N_s} - \mathcal{E}_\mathrm{VMC}^{N_s},
 \label{eq:1b_vmc}
\end{equation}
where $\mathcal{E}_\mathrm{VMC}^{N_s}$ is the EST energy in a $N_s$-atom supercell, 
whereas $\mathcal{E}_\mathrm{VMC}^{\mathrm{TABC},N_s}$ is the correspondent fully converged 
TABC result in the same supercell. Results for Li and Si
are presented in Tab.~\ref{tab:2b_errors} ($4$th column); they constitute an indirect probe of the 
changes in the Fermi surface when going from DFT to QMC level. If one takes bcc-Li as the reference
case, one can see that the one-body residual corrections are very small (one order of magnitude smaller than
many-body effects) and decrease fast when the size is increased. This implies that, in this case, 
the estimated value of the EST is supposedly very close to the one obtained at the DFT(LDA) level. This is further
confirmed by the fact that the EST value changes only slightly from simple Hartree to DFT(LDA) mean-field 
estimates. In other words, the actual EST is rather insensitive to 
the underlying theory used to determine it which makes the EST evaluation quite robust. 

Moreover, we notice that for $\beta$-tin Si the behavior of $\epsilon_{\text{1b}}^{\mathrm{VMC}}$ is less 
systematic as a function of the system size. This could be related to the limitation of the TABC approach
used to estimate the one-body corrections at VMC level. Indeed, as already mentioned, TABC works in the 
canonical ensemble and can introduce a bias in the energy values due to the wrong $\mathbf{k}$-point occupations,
that can be particularly severe in the case of $\beta$-tin Si where the Fermi
surface is much more complex than in the Li case (see 
Fig.~\ref{fig:fermi}).

The residual many-body contribution to the FS errors can also be evaluated at VMC level. 
In Tab.~\ref{tab:2b_errors} we compare the many-body errors estimations
obtained with the standard KZK method in Eq.~\ref{eq:kzk} ($2$nd column) and directly within VMC ($3$rd colum) 
using the relation:
\begin{equation}
 \epsilon_{\text{2b}}^{\mathrm{VMC}} = \mathcal{E}_\mathrm{VMC}^{\text{extr}} - \mathcal{E}_\mathrm{VMC}^\mathrm{TABC,N_s} 
 \label{eq:2b_vmc}
\end{equation}
where $\mathcal{E}_\mathrm{VMC}^{\text{extr}}$ is the VMC energy extrapolated to the infinite-size limit.
Notice that this extrapolation is obtained using KZK corrected values, however the infinite size limit 
must obviously be the same. 

Tab.~\ref{tab:2b_errors} demonstrate that KZK and direct VMC estimations are in 
good agreement concerning many-body FS errors for both Li and Si, thus supporting the use 
of the cheap KZK approach for production QMC runs. 
The only relevant discrepancy is shown by the Si 16 atoms supercell
which is a particularly delicate case both for EST and TABC methods, as previously mentioned.

In conclusion, it is important to remark that the special twist value used for Li and Si calculations has been 
determined using DFT(LDA) energies, therefore its value can be
slightly different than the one
obtained directly with the DFT(KZK) functional instead. 
This discrepancy could lead to some spurious contribution to the KZK estimation
of the many-body errors. However, we verified that the variation of the EST 
between the two functionals is negligible, in line with what said before, thus 
$\epsilon_{\text{2b}}^{\mathrm{KZK}}$ is considered as purely many-body and directly comparable 
with $\epsilon_{\text{2b}}^{\mathrm{VMC}}$.

\subsubsection{Energy derivatives}
\label{sec:derivatives}

\begin{table}[!htp]
\begin{center}
\caption { Lithium cell parameters obtained from VMC structural relaxation at different supercell sizes. The EST
           results are compared with standard PBC calculations and with the most accurate TABC method. 
           We report also values obtained from fully converged DFT calculations
           in the LDA approximation, performed with the QuantumESPRESSO\cite{qe} program using 
           a $15\times15\times15$ $\mathbf{k}$-point mesh and norm conserving pseudopotentials. 
           Experimental cell parameters are also shown. } \label{tab:geoLi}
           \vspace{0.2cm}
\begin{tabular}{c|ccc} 

   \hline
          
   $\#$ \textrm{of atoms} & \multicolumn{3}{c}{ \textrm{Cell parameter} [{\AA}] } \\
       & PBC & EST & TABC \\
   \hline

    16       &   3.497(6)  &  3.457(4)  &  3.454(4) \\
    54       &   3.469(3)  &  3.476(3)  &  3.496(3) \\
    128      &   3.521(3)  &  3.505(3)  &  3.502(2) \\
    250      &   3.510(5)  &  3.506(2)  &  3.499(3) \\
    DFT(LDA) &   3.3537    &            &  \\
    EXP\cite{lithium_exp} &  3.482 &   &  \\   
    
  \hline 
  
\end{tabular}
\end{center}
\end{table} 

\begin{figure}
     \caption{ Hydrogen radial pair distribution function extracted from a molecular dynamics simulation 
     of liquid hydrogen at a density given by $r_s = 1.36$ a.u. 
     For this plot we use a supercell of 64 atoms. We compare our EST method with the standard
     PBC, with the Baldereschi point and with TABC results performed 
     with a $4\times4\times4$ uniform mesh (64 twists). The solid lines are
     obtained with a polynomial interpolation as guide for the eye. 
     Despite small discrepancies in the $g(r)$ around the peaks, 
     TABC and EST display an overall good agreement.
     We notice instead some spurious features in the results obtained with the Baldereschi point,
     in particular in the region zoomed in the inset ($r \in [1.5,3.2]$). } \label{fig:corrFuncs64}
     \includegraphics[width=\columnwidth]{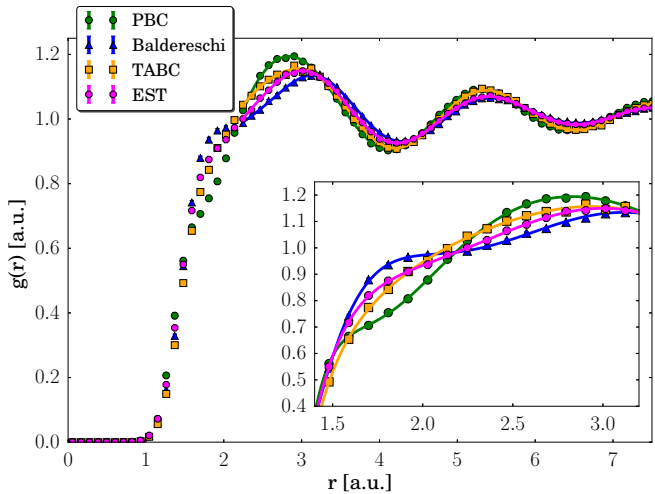}
\end{figure}

\begin{figure}
     \caption{ Same as Fig.~\ref{fig:corrFuncs64} but performed with a larger supercell containing 128 hydrogen atoms. 
     In this case the disorder of the system prevails and the system can be considered spherically symmetric; for this
     reason the special twist coincides with the Baldereschi point.
     We notice that at this supercell size the EST and TABC curves 
     are practically indistinguishable, while PBC is still far from the 
     other two methods, especially in the description of the first peak around $r=3.2$ a.u.} \label{fig:corrFuncs128}
     \includegraphics[width=\columnwidth]{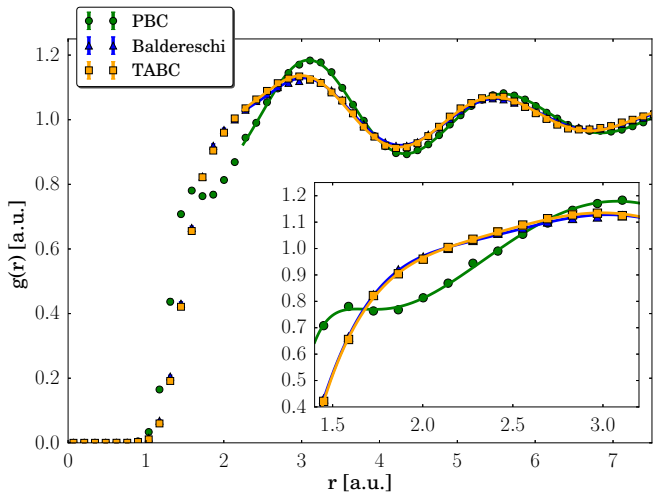}
\end{figure}

In Sec.~\ref{sec:energetics} we demonstrated the reliability of the EST method in extrapolating QMC energies to
the infinite-size limit.
However, 
as already pointed out in Ref.~\onlinecite{ceperley_twist_average}, sampling the Fermi surface 
with a single point might not 
be sufficient to account for more sensitive properties of the system such as the potential energy or 
correlation functions. 
In this Section we focus on testing the EST method with a particularly important type of correlation function: 
the ionic forces. Their evaluation within QMC has been the subject of intense research 
activity\cite{corrSampl_forces,sorella_aad} due to the intrinsic 
difficulty to find an efficient and finite variance algorithm for computing many-body energy derivatives.
As already mentioned, the AAD technique\cite{sorella_aad} offers a solution to this issue.
With our approach is thus possible to perform both zero temperature structural relaxation 
and molecular dynamics simulations based on QMC forces\cite{MDhydro,MDhydro2,MDwater}.

In Tab.~\ref{tab:geoLi} we present the optimization of the cell parameter of bcc Li carried out with full 
QMC forces minimization.
The EST results are obtained with the same value of special twist 
previously used for energy extrapolation.

Thanks to its denser sampling of the Fermi surface, the TABC method
performs better than single-twist methods, with the cell parameter
already converged for the 54 atoms supercell.  
EST is slightly slower to converge towards the infinite-size limit than TABC, but it displays
a much smoother behavior with respect to simple PBC calculations, thus allowing an easy extrapolation
to infinite size. 
By performing a linear extrapolation we obtain
3.508(3) \AA\ for TABC and 3.504(5) \AA\ for EST, which are in statistical agreement. The residual discrepancy
with the experimental value\cite{lithium_exp} ($\sim 0.025$~\AA) can be due to temperature effects which
are not taken into account by our calculations.
However, QMC outcome provides already a substantial improvement with respect to DFT(LDA) calculations. 

The lower computational cost of the EST method makes this approach the
appropriate choice in the case of structural relaxation of more complex crystal cells, 
requiring the use of large supercells or 
the simultaneous optimization of several structural parameters, 
which would be computationally infeasible using the TABC technique.

The last part of this Section is devoted to benchmark calculations on the radial pair 
distribution function ($g(r)$) of liquid
hydrogen. The $g(r)$ is extracted from QMC based molecular dynamics (MD) simulations at a temperature of $1800$ K and
an estimated pressure of $\simeq 260$ GPa.
Forces are computed with the AAD 
technique and the MD is carried out with the methods introduced in Refs.~\onlinecite{MDhydro,MDhydro2}.
At these conditions, in our simulations liquid hydrogen is in the atomic phase\cite{MDhydro}. This phase is metallic, 
hence FS effects are likely to be important in order to obtain a reliable description of the system.

The evaluation of the special twist for liquid hydrogen MD is more challenging. 
Due to its disordered nature, the system can be considered spherically symmetric and thus one can assume
that the Baldereschi point for cubic systems $(1/4,1/4,1/4)$ would provide a good approximation for the 
special twist, since the disorder favors the lowest Fourier components of Brillouin zone integrals.
In order to investigate better this issue, 
we extract several configurations from a previous molecular dynamics simulation carried out with the 
same conditions and we apply the EST procedure. The special twist is obtained by averaging among 
all the points found. In the case of the $64$ atoms supercell, 
we realized that the special twist has no relevant fluctuations
among different configurations and that it tends to a value of $(1/4,1/4,0)$, in contrast with the initial
assumption; we conclude that within this relatively small supercell the system tends to break the 
spherical (cubic) symmetry given by its disordered nature. 
On the contrary, for the $128$ supercell we found a special twist very close 
the cubic Baldereschi point, as expected. This demonstrates that this supercell is sufficient to correctly
account for the disordered nature of liquid hydrogen.

In Fig.~\ref{fig:corrFuncs64} the $g(r)$ is reported for the $64$ atoms supercell. 
The overall behavior of the EST curve is in good
agreement with TABC with only small discrepancies around the peaks. 
This is not the case for simulations carried out with the Baldereschi point, where 
we notice an anomalous feature close to the first peak at $r \sim 2.8$ a.u. 
An excellent agreement between the two methods is instead obtained with a $128$ atoms supercell, as shown in
Fig.~\ref{fig:corrFuncs128} where TABC and EST (Baldereschi) curves perfectly superpose.
At difference with the EST method, also with this larger supercell the $g(r)$ extracted from PBC calculations 
qualitatively differs from the more accurate TABC result.

The EST method remains reliable in the evaluation of 
ionic forces, in both simple structural relaxation and pair distribution functions extracted from MD simulations. However, 
the accordance with the more expensive TABC technique is achieved only when relatively large supercells are 
employed. Hence, if one has to compute correlations functions using the EST method, a 
careful assessment of the impact of supercell size is needed before starting production runs.

\section{Conclusions} \label{sec:conclusions}

In this manuscript we presented a novel procedure, 
dubbed EST method, to find special twist values in the 
Brillouin zone corresponding to the simulation cell, which reproduce the mean-field infinite-size energy
up to an arbitrarily high numerical accuracy.
We show that, when the EST value is used to build wavefunctions for correlated \emph{ab initio} QMC calculations,
it greatly reduces one-body shell fluctuations in the energy
extrapolation to the infinite-size limit.

Our procedure has several advantages with respect to the widely used TABC technique. From a computational point of view, 
it is a single-twist technique and it is therefore considerably more affordable, especially within diffusion Monte Carlo 
calculations characterized by a significant equilibration time. Within our method, it is not only possible
to accurately compute thermodynamic converged total energies, but its relatively low cost also ensures
the possibility of performing structural relaxation of complex supercells or even large molecular dynamics 
simulations within the QMC framework. 
On the other hand, the EST method is constructed in order to keep the exact mean-field 
thermodynamic limit of the many-body variational wavefunction. This feature allows to 
avoid any bias in the kinetic energy evaluation and provides a more reliable description of the Fermi surface
when a large supercell is used. 

We tested our procedure on the 3D electron gas, a simple, but widely studied model for metallic correlated systems. 
Within this system EST displays an efficiency comparable with the standard TABC technique. 
We demonstrate that EST is also very effective in controlling FS effects when tackling more complex 
and realistic systems, such as solid hydrogen, bcc-Li and the high-pressure $\beta$-tin phase of silicon. 
These systems show different degrees of complexity and they represent an exhaustive testing ground for our method. 
$\beta$-tin Si is particularly challenging, since it possesses a very complicated Fermi surface. EST and TABC
extrapolated results are shown in very good agreement, and the two methods ensure a very similar smoothing of the 
one-body energy fluctuations. 

The calculation of correlation functions such as ionic forces is more delicate. 
We show for both zero temperature structural relaxation and molecular dynamics simulations 
that EST performs better than any other single-twist method. However, the TABC technique 
still shows a better performance, 
thanks to its denser sampling of the Fermi surface.
We show that for reasonably large supercell sizes, EST and TABC techniques are in perfect agreement. Therefore
a careful study of the supercell size dependence is necessary before
applying the EST method in QMC production runs for the calculation of correlation functions.

We believe that EST procedure here introduced can be the method of choice for reducing FS effects in many practical situations,
particularly when the complexity of the system or the required supercell size make the more demanding
TABC calculations infeasible. Last but not least, given its single-twist nature, EST can also be efficiently 
used in combination with full determinant optimization of the wavefunction, 
QMC structural relaxation and molecular dynamics simulations, as we
demonstrated in this paper.

\acknowledgements
M. D. would like to thank Ruben Weht for useful discussions. 
We acknowledge computational resources provided through the HPCI
System Research Project No. hp140092 on the K computer
at RIKEN Advanced Institute for Computational Science, and on the
HOKUSAI GreatWave computer under the project G16026. 
M. C. thanks the GENCI French program to provide additional computer
time through the Grant No. 096493.
 
\bibliography{manuscript_bibliography}

\end{document}